\documentclass[preprint]{JHEP}
\usepackage{epsfig}
\usepackage{showkeys}
\def\be{\begin{equation}}
\def\ee{\end{equation}}
\def\bs{\begin{subequations}}
\def\es{\end{subequations}}
\newcommand{\een}{\end{subequations}}
\newcommand{\ben}{\begin{subequations}}
\newcommand{\beq}{\begin{eqalignno}}
\newcommand{\eeq}{\end{eqalignno}}

\def\l{\lambda}
\def\m{\mu}

\def\sp{\;\;\;,\;\;\;}
\def\r{\rho}

\def\beq{\begin{equation}}
\def\eeq{\end{equation}}

\def\4R{{{}^{(4)}R}}

\def\K5{{\kappa}}
\def\K52{{\kappa^2}}

\newcommand{\drho}{\delta \rho}

\newcommand{\dchi}{\delta \chi}

\newcommand{\ii}{i}
\newcommand{\jj}{j}

\newcommand{\da}{\dot{a}}
\newcommand{\db}{\dot{b}}
\newcommand{\dn}{\dot{n}}
\newcommand{\dda}{\ddot{a}}
\newcommand{\ddb}{\ddot{b}}

\newcommand{\pa}{a^{\prime}}
\newcommand{\pb}{b^{\prime}}
\newcommand{\pn}{n^{\prime}}
\newcommand{\ppa}{a^{\prime \prime}}

\newcommand{\ppn}{n^{\prime \prime}}
\newcommand{\fda}{\frac{\da}{a}}
\newcommand{\fdb}{\frac{\db}{b}}
\newcommand{\fdn}{\frac{\dn}{n}}
\newcommand{\fdda}{\frac{\dda}{a}}
\newcommand{\fddb}{\frac{\ddb}{b}}

\newcommand{\fpa}{\frac{\pa}{a}}
\newcommand{\fpb}{\frac{\pb}{b}}
\newcommand{\fpn}{\frac{\pn}{n}}
\newcommand{\fppa}{\frac{\ppa}{a}}

\newcommand{\fppn}{\frac{\ppn}{n}}

\def\bea{\begin{eqnarray}}                     
\def\eea{\end{eqnarray}}                       
\title{Brane-bulk energy exchange and cosmological acceleration   }
\author{
Elias Kiritsis\\
CPHT, Ecole Polytechnique,\\
 91128, Palaiseau, FRANCE\\
  UMR du CNRS 7644.\\
and\\
Department of Physics, University of Crete\\
71003 Heraklion, GREECE\\
{\tt E-mail from http://cpht.polytechnique.fr/cpht/kiritsis/} }

\preprint{\hepth{0503189} \\ CPHT RR 037.0704}

\abstract{The consequences for the brane cosmological evolution of energy exchange
between the brane and the bulk
are analyzed.
A rich variety of brane cosmologies is obtained, depending on
the precise mechanism of energy transfer,
the equation of state of brane-matter
and the spatial topology. An accelerating era is generically a feature
of the solutions.
\\
\phantom{.}
\\
\phantom{.}
\\
\phantom{.}
\\
\phantom{.}
\\
\tt Prepared for 36th International Symposium Ahrenshoop on the Theory 
of Elementary Particles: Recent Developments in String M Theory and
 Field Theory, Wernsdorf, Germany, 26-30 Aug 2003.\\

\bf Published in Fortsch.Phys.52:568-577,2004

}

\begin{document}

\section{Introduction}

My review lecture  has focused on some recent progress in
the physics of brane-worlds. Both  model
building and  cosmological implications were considered.
Since there is already a published review on such issues
\cite{review}, and since space here is scarce, I have decided to
develop a particular aspect of brane cosmology, which is both
important and little studied, namely that of brane-bulk energy
exchange. I will review recent progress and present some new
results. There has been some study of cosmological aspects of brane-bulk energy
exchange in the recent past \cite{ex1}-\cite{ex4}.
Here, I will mostly focus on the approach described in \cite{kkttz}.

In any theory, where gravity is higher-dimensional classically
(around flat space), a mechanism must be invoked to explain why
observable gravity is four-dimensional at experimentally
verifiable distances. The easiest and most popular mechanism is
compactification, but alternatives have been also considered
lately, namely RS localisation (4d in the IR) \cite{rs} and
brane-induced gravity \cite{ig} (4d in the UV).
In all such realizations, the 4d graviton is accompanied by KK
modes that propagate in the bulk and can thus mediate
energy-exchanging processes between brane and bulk matter.
Such processes are important because on the one hand they provide
stringent constraints on brane-world models and on the other as it
was argued in \cite{kkttz} they may generate interesting
cosmological effects.

In \cite{ktt} the cosmological brane-bulk energy exchange
processes , due to KK gravitons were studied in the simplest RS cosmological
evolution supplemented with a four-dimensional (induced) Einstein term on the brane .
It was shown that, in agreement with intuition,  such processes do
not affect the cosmological evolution during 4d eras ($H\sim
\sqrt{\rho}$), while they are important in higher-dimensional eras
($H\sim \rho$).
Moreover, in many important cases, the rate of energy loss can be
written as a power of the driving brane-energy density, something
that can be used in more general situations.

To generalize, we consider  more bulk fields and more general bulk-brane couplings.
Although this can be formulated by the standard action principle and the relevant exact equations
derived and studied we will take a short-cut. We will analyze the regime where the bulk energy, at the brane position,
can be consistently neglected from the equations. Moreover,
we parameterize appropriately the energy exchange as a specific power of
the matter density of the brane.

The effects found, include accelerated evolution with standard
$w=0$ matter and outflow, inflationary fixed points in the
presence of inflow, and tracking behavior between observable and
mirage (dark) energy.

\section{The model}

We shall be interested in the model described by the action
\be
S=\int d^5x~ \sqrt{-g} \left( M^3 R -\Lambda +{\cal L}_B^{mat}\right)
+\int d^4 x\sqrt{-\hat g} \,\left( -V+{\cal L}_b^{mat} \right),
\label{001}
\ee
where $R$ is the curvature scalar of the five-dimensional metric
$g_{AB}, A,B=0,1,2,3,5$,
$\Lambda$ is the bulk cosmological constant, and
${\hat g}_{\alpha \beta}$, with $\alpha,\beta=0,1,2,3$,
is the induced metric on the 3-brane.
We identify
$(x,z)$ with $(x,-z)$, where $z\equiv x_5$. However, following the conventions
of \cite{rs} we extend the bulk integration over the entire interval
$(-\infty,\infty)$.
The quantity $V$ includes the brane tension as well as
quantum contributions to the
four-dimensional cosmological constant.

We consider an ansatz for the metric of the form
\begin{equation}
ds^{2}=-n^{2}(t,z) dt^{2}+a^{2}(t,z)\gamma_{ij}dx^{i}dx^{j}
+b^{2}(t,z)dz^{2},
\label{metric}
\end{equation}
where $\gamma_{ij}$ is a maximally symmetric 3-dimensional metric.
We use $\tilde k$ to parameterize the spatial curvature.

The non-zero components of the five-dimensional Einstein tensor are
\begin{eqnarray}
{G}_{00} &=& 3\left\{ \fda \left( \fda+ \fdb \right) - \frac{n^2}{b^2}
\left(\fppa + \fpa \left( \fpa - \fpb \right) \right) + \tilde k \frac{n^2}{a^2} \right\},
\label{ein00} \\
 {G}_{\ii\jj} &=&
\frac{a^2}{b^2} \gamma_{ij}\left\{\fpa
\left(\fpa+2\fpn\right)-\fpb\left(\fpn+2\fpa\right)
+2\fppa+\fppn\right\}
\nonumber \\
& &+\frac{a^2}{n^2} \gamma_{ij} \left\{ \fda \left(-\fda+2\fdn\right)-2\fdda
+ \fdb \left(-2\fda + \fdn \right) - \fddb \right\} -\tilde k \gamma_{ij},
\label{einij} \\
{G}_{05} &=&  3\left(\fpn \fda + \fpa \fdb - \frac{\dot{a}^{\prime}}{a}
 \right),
\label{ein05} \\
{G}_{55} &=& 3\left\{ \fpa \left(\fpa+\fpn \right) - \frac{b^2}{n^2}
\left(\fda \left(\fda-\fdn \right) + \fdda\right) - \tilde k \frac{b^2}{a^2}\right\}.
\label{ein55}
\end{eqnarray}
Primes indicate derivatives with respect to
$z$, while dots derivatives with respect to $t$.

The five-dimensional Einstein equations take the usual form
\beq
G_{AC}
= \frac{1}{2 M^3} T_{AC} \;,
\label{einstein}
\eeq
where $T_{AC}$ denotes the total energy-momentum tensor.

Assuming a perfect fluid on the brane and, possibly an additional energy-momentum
$T^A_C|_{m,B}$ in the bulk, we write
\begin{eqnarray}
T^A_{~C}&=&
\left. T^A_{~C}\right|_{{\rm v},b}
+\left. T^A_{~C}\right|_{m,b}
+\left. T^A_{~C}\right|_{{\rm v},B}
+\left. T^A_{~C}\right|_{m,B}
\label{tmn1} \\
\left. T^A_{~C}\right|_{{\rm v},b}&=&
\frac{\delta(z)}{b}{\rm diag}(-V,-V,-V,-V,0)
\label{tmn2} \\
\left. T^A_{~C}\right|_{{\rm v},B}&=&
{\rm diag}(-\Lambda,-\Lambda,-\Lambda,-\Lambda,-\Lambda)
\label{tmn3} \\
\left. T^A_{~C}\right|_{{\rm m},b}&=&
\frac{\delta(z)}{b}{\rm diag}(-\tilde\rho,\tilde p,\tilde p,\tilde p,0),
\label{tmn4}
\end{eqnarray}
where $\tilde \rho$ and $\tilde p$ are the energy density and pressure on the brane, respectively.
The
behavior of $T^A_C|_{m,B}$ is in general complicated in the presence
of flows, but we will not  specify it further at this point.

We wish to solve the Einstein equations at the location
of the brane following \cite{binetruy}. We indicate by the subscript $o$ the value of
various quantities on the brane.
Integrating equations (\ref{ein00}), (\ref{einij})
with respect to $z$ around $z=0$ gives the known
jump conditions
\begin{eqnarray}
a_{o^+}'=-a_{o^-}'  &=& -\frac{1}{12M^3} b_o a_o \left( V +\tilde \rho \right)
\label{ap0} \\
n'_{o^+}=-n_{o^-}' &=&  \frac{1}{12M^3} b_o n_o \left(- V +2\tilde \rho +3 \tilde p\right).
\label{np0}
\end{eqnarray}

The other two Einstein equations (\ref{ein05}), (\ref{ein55})
give
\begin{equation}
\frac{n'_o}{n_o}\frac{\dot a_o}{a_o}
+\frac{a'_o}{a_o}\frac{\dot b_o}{b_o}
-\frac{\dot a'_o}{a_o} =
\frac{1}{6M^3}T_{05}
\label{la1}
\end{equation}
\begin{equation}
\frac{a'_o}{a_o}\left(
\frac{a'_o}{a_o}+\frac{n'_o}{n_o}\right)
-\frac{b^2_o}{n^2_o}\left(
\frac{\dot a_o}{a_o} \left( \frac{\dot a_o}{a_o}-\frac{\dot n_o}{n_o}\right)
+\frac{\ddot a_o}{a_o}\right)
-\tilde k\frac{b^2_o}{a^2_o} =-\frac{1}{6M^3}\Lambda b^2_o
+ \frac{1}{6M^3}T_{55},
\label{la2}
\end{equation}
where $T_{05}, T_{55}$ are the $05$ and $55$ components of $T_{AC}|_{m,B}$
evaluated on the brane.
Substituting (\ref{ap0}), (\ref{np0})
in equations (\ref{la1}), (\ref{la2}) one obtains
\begin{equation}
\dot {\tilde\rho} + 3 \frac{\dot a_o}{a_o} (\tilde\rho +\tilde p)
= -\frac{2n^2_o}{b_o}
T^0_{~5}
\label{la3}
\end{equation}
\begin{eqnarray}
\frac{1}{n^2_o} \Biggl(
\frac{\ddot a_o}{a_o}
+\left( \frac{\dot a_o}{a_o} \right)^2
&-&\frac{\dot a_o}{a_o}\frac{\dot n_o}{n_o}\Biggr)
+\frac{\tilde k}{a^2_o}
=\frac{1}{6M^3} \Bigl(\Lambda + \frac{1}{12M^3} V^2
\Bigr)
\nonumber \\
&-&\frac{1}{144 M^6} \left(
V (3\tilde p-\tilde \rho ) +\tilde \rho (3\tilde p +\rho)
\right)
- \frac{1}{6M^3}T^5_{~5}.
\label{la4}
\end{eqnarray}

In the model that reduces to the Randall-Sundrum
vacuum \cite{rs} in the absence of matter, the first
term on the right hand side of equation (\ref{la4}) vanishes. A new scale $K$
may be  defined through the relations
$V=-\Lambda/K=12M^3 K$.
Sometimes, we will relax this condition.

At this point we find it convenient to employ a coordinate frame in
which $b_o=n_o=1$ in the above equations. This can be achieved by using Gauss
normal coordinates with $b(t,z)=1$, and by going to the temporal gauge on the
brane with $n_o=1.$ The assumptions for the form of the energy-momentum
tensor are then specific to this frame.
Using $\beta\equiv M^{-6}/144$ and $\gamma\equiv V M^{-6}/144$,
and omitting the subscript o for convenience in the following, we rewrite
equations (\ref{la3}) and (\ref{la4}) in the equivalent first order form
\be
\dot{\tilde\rho}+3(1+w)\,{{\dot a}\over a} \,\tilde \rho = -\tilde
T\sp
{{{\dot a}^2}\over {a^2}}=\beta\tilde\rho^2+2\gamma (\tilde\rho+\tilde\chi) -
{\tilde k\over{a^2}}+\tilde \lambda
\label{aa}
\ee
\be
\dot{\tilde\chi}+4\,{{\dot a}\over
a}\,\tilde\chi=\left({\beta\over \gamma}\tilde \rho+1\right)\tilde T-{1\over 6\gamma~M^3}{{\dot a}\over
a}{T^5}_5,
\label{chi}
\ee
where $\tilde p=w\tilde \rho$,
$\tilde T=2T^0_{~5}$ is the discontinuity of the zero-five component of the bulk
energy-momentum tensor,
and $\tilde \lambda=(\Lambda+V^2/12M^3)/12M^3$ the effective
cosmological constant on the brane.

In the equations above, Eq. (\ref{aa}) is the $definition$ of the auxiliary density $\tilde\chi$.
With this definition, the other two equations are equivalent to the original system
(\ref{la3},\ref{la4}).
As we will see later on, in the special case of no-exchange ($\tilde T=0$) $\tilde \chi$ represents
the mirage radiation reflecting the non-zero Weyl tensor of the bulk.

The second order equation (\ref{la4}) for the scale factor becomes
\be
{{\ddot a}\over a}=-(2+3w)\beta\tilde\rho^2-(1+3w)\gamma\tilde \rho-2\gamma\tilde \chi+\tilde \lambda.
\label{decel}
\ee
As mentioned above, in the Randall-Sundrum model
the effective cosmological constant $\tilde \lambda$ vanishes, and this is the value we shall
assume for most of the rest.

We may now pass to dimensionless quantities.
The AdS scale is defined as  $V=M^3 K$ and the dimensionless
densities as
\be
\rho={\tilde\rho\over 72 M^3K}\sp  \chi={\tilde\chi\over 72
M^3K}\sp \lambda ={\tilde\lambda \over K^2}={\Lambda\over 12M^3K^2}+{1\over 144}
\ee
\be
T={\tilde T\over 72 M^3K^2}\sp T_5={{ T^5}_5\over 3M^3K^2}
\ee
as well as a rescaled time $\tau =K~t$.
The combination $k=\tilde k/K^2$ is then dimensionless and by scaling $a$ we can set it to
$ k=0,\pm 1$.
Then the cosmological equations
become
\be
\dot{\rho}+3(1+w)\,{{\dot a}\over a} \, \rho = -T
\sp
{{{\dot a}^2}\over {a^2}}=36\rho^2+\rho+\chi -
{k\over{a^2}}+\lambda
\label{rho1}\ee
\be
\dot\chi+4\,{{\dot a}\over
a}\,\chi=\left(72\rho+1\right) T-{{\dot a}\over
a}T_5,\sp
q\equiv {\ddot a\over a}=-36(2+3w) \rho^2-{3w+1\over
2}\rho-\chi-{1\over 2}T_5+\lambda
\label{qq}\ee
where now dots stand for derivatives with respect to $\tau$.

We may separate the dynamics associated with ${T^0}_5$ and
${T^5}_5$ by introducing a new energy density $\sigma$ so that
\be
{{{\dot a}^2}\over {a^2}}=36\rho^2+\rho+\chi +\sigma-
{k\over{a^2}}+\lambda\sp \dot\chi+4\,{{\dot a}\over a}\,\chi=\left(72\rho+1\right) T\sp \dot\sigma+4\,{{\dot a}\over
a}\,\sigma=-{{\dot a}\over
a}T_5
\ee

In order to derive a solution
that is largely independent of the bulk dynamics,
the $T^5_{~5}$ term on the
right hand side of the same equation must be
negligible relative to the second one.
This is possible if we assume that the diagonal elements of the various
contributions to the energy-momentum tensor satisfy the schematic
inequality \cite{kkttz}
\be
\left|
\frac{\left. T\right|^{\rm diag}_{{\rm m},B}}{
\left. T\right|^{\rm diag}_{{\rm v},B}}
\right|
\ll
\left|
\frac{\left. T\right|^{\rm diag}_{{\rm m},b}}{
\left. T\right|^{\rm diag}_{{\rm v},b}}
\right|.
\label{vacdom} \ee
Our assumption is that the bulk matter
relative to the bulk vacuum energy is much less important than
the brane matter relative to the brane vacuum energy. In this case
the bulk is largely unperturbed by the exchange of energy with the brane.
When the off-diagonal term $T^0_{~5}$ is of the same order of magnitude or
smaller than the diagonal ones, the inequality (\ref{vacdom}) implies
$T\ll \tilde\rho K$.

Thus, under this assumption, we may neglect $T_5$ from
(\ref{qq}).
Moreover, we will parameterize $T$ as a function of the driving energy density $\rho$.
The precise form depends on the details of the interaction between
brane and bulk. In the AdS scaling region this has typically a
power dependence $T\sim \rho^{\nu}$.

\subsection{Other forms of the cosmological equations}.

Combining equations (\ref{rho1}) we obtain
\be
a{{d\rho}\over da}=-3(1+w)\rho-
\epsilon\,T(\rho) \left(36\rho^2+\rho+\chi -
{ k\over{a^2}}+\lambda\right)^{-1/2}.
\label{rho(a2)}
\ee
Similarly, equations (\ref{rho1}) and (\ref{qq}) give
\be
a{{d\chi}\over da}=-4\chi+\epsilon \left(36\rho+1\right)
T(\rho)
\left(36\rho^2+\rho+\chi -
{k\over{a^2}}+\lambda\right)^{-1/2},
\label{chi(a2)}
\ee
where $\epsilon =1$ refers to expansion, while $\epsilon=-1$ to contraction.
These two equations form a two-dimensional dynamical system. The function
$\chi(\rho)$ is obtained from the equation
\begin{eqnarray}
\Biggl(3(1+w)\rho
\sqrt{36\rho^2+\rho+\chi -
{k\over{a^2}}+\lambda}&+&\epsilon\,T(\rho)
\Biggr)\,
{{d\chi}\over{d\rho}}
\nonumber \\
=4\chi\sqrt{36\rho^2+\rho+\chi -
{k\over{a^2}}+\lambda}
&-&\epsilon \left(36\rho+1\right) T(\rho).
\label{chi(rho)}
\end{eqnarray}
Note that the equations of contraction are those of expansion with the
roles of outflow and influx interchanged.

For $k=0$, $\hat T=
\rho^{-{3/2}} T$ and $\zeta=\sqrt{36 \rho
+{\chi\over\rho}+1+{\lambda\over \rho}}$ the equation can be
simplified
\be
2\rho\zeta\zeta'={(1-3w)(\zeta^3-\zeta)+72(1+3w)\rho\zeta-\hat
T\zeta^2+36\rho\hat T-4{\lambda \zeta\over \rho}\over
3(1+w)\zeta+\hat T}
\ee

If we further define
\be
\zeta=\rho^{1-3w\over 6(1+w)}\xi
\ee
we obtain
\be
2\rho\xi\xi'={-\left({\hat T\over 3(1+w)}\right)\rho^{1-3w\over
3(1+w)}\xi^2+\left[3w-1+72(1+3w)\rho-4{\lambda\over
\rho}\right]\rho^{1-3w\over 6(1+w)}\xi+36\rho\hat T \over 3(1+w)\rho^{1-3w\over 6(1+w)}\xi+\hat T}
\ee

Finally the  acceleration satisfies
\be
-{dq\over d\rho}={\left[72(1+3w)(2+3w)\rho^2+{9w^2-1\over
2}\rho+4\lambda-4q\right]R+108(2w+1)\rho T+{3w-1\over 2}T\over
3(1+w)\rho R+T}
\ee
where
\be
R=\sqrt{{1-3w\over 2}\rho-36(1+3w)\rho^2+2\lambda-q-{k\over a^2}}
\ee

\subsection{The four-dimensional regime}
When  $\rho<<1$ we are in the 4d regime ($H^2\sim \rho$). Here the
cosmological equations simplify

\be
\dot{\rho}+3(1+w)\,{{\dot a}\over a} \, \rho = -T\sp
{{{\dot a}^2}\over {a^2}}=\rho+\chi -
{k\over{a^2}}+\lambda
\label{a11}
\ee
\be
\dot\chi+4\,{{\dot a}\over
a}\,\chi=T
\sp
q={\ddot a\over a}=-{3w+1\over
2}\rho-\chi+\lambda
\label{aa11}\ee

For $k=0$, $\hat T=
\rho^{-{3/2}} T$ and $\zeta=\sqrt{{\chi\over\rho}+1+{\lambda\over \rho}}$
Then
\be
2\rho\zeta'={(1-3w)(\zeta^2-1)-\hat
T\zeta-4{\lambda\over \rho}\over
3(1+w)\zeta+\hat T}
\label{zz}
\ee
while (\ref{a11}) becomes
\be
a{d\log \rho\over da}=-3(1+w)-{\hat T\over \zeta}
\label{zzz}\ee

\section{Inflating fixed points}

An interesting feature of the cosmological equations is the possible presence of
accelerating cosmological solutions. We may look for
exponential expansion with a constant Hubble parameter $H$, even if the
brane content is not pure vacuum energy.
We will restrict ourselves for simplicity to the 4d regime. For
the non-linear analysis we refer the reader to \cite{kkttz}.

For a  fixed point
equations (\ref{a11}), (\ref{aa11}) must have a time-independent
solution, without necessarily requiring $w=-1$ ($\lambda=k=0$).
The possible fixed points (denoted by $*$) of these equations for
$k=0$ satisfy
\begin{eqnarray}
3H_*(1+w)\rho_* = -T(\rho_*)
\sp
H^2_* =  \rho_* + \chi_*
\sp
4 H_* \chi_* =  T(\rho_*).
\label{fp}
\end{eqnarray}
It is clear from equation (\ref{fp}) that, for positive matter density
on the brane ($\rho >0$), flow of energy into the brane
($T(\rho)<0$) is necessary.

The accretion of energy from the bulk depends on the
dynamical mechanism that localizes particles on the brane.
Its details are outside the scope of our discussion. However, it is
not difficult to imagine scenaria that would lead to accretion.
If the brane initially has very low energy density,
energy can by transferred onto it by
bulk particles such as gravitons.
An equilibrium is expected to set in
if the brane energy density reaches a limiting value. As a result,
a physically  motivated behavior for the function
$T(\rho)$ is to be negative for small $\rho$ and cross zero towards positive
values for larger densities.
In the case of accretion it is also natural to expect that the energy
transfer approaches a negative constant value for $\rho \to 0$.

The solution of equations (\ref{fp})
satisfies
\begin{eqnarray}
T(\rho_*) &=& -\frac{3}{2} (1+w)\sqrt{1-3w}~\rho_*^{3/2}
\sp
H_*^2 = \frac{1-3w}{4}  \rho_*
\sp
\chi_* = - \frac{3(1+w)}{4}  \rho_*.
\label{sol}\end{eqnarray}
For a general form of
$T(\rho)$ equation (\ref{sol}) is an algebraic equation with
a discrete number of roots.
For any value of $w$ in the region
$-1<w<1/3$ a solution is possible.
The corresponding cosmological model has a scale factor that
grows exponentially with time. The energy density on the brane remains
constant due to the energy flow from the bulk.
This is very similar to the steady state model of cosmology
\cite{steady}. The main
differences are that the energy density is not spontaneously
generated, and the Hubble parameter receives an additional contribution from
the ``mirage'' field $\chi$ (see equation (\ref{fp})).

The stability of the fixed point
(\ref{fp}) determines whether the exponentially expanding
solution is an attractor of neighboring cosmological flows. If we
consider a small homogeneous perturbation around the fixed point
($\rho=\rho_*+\drho$, $\chi=\chi_*+\dchi$) we find that $\drho,\dchi$ satisfy
\be
\frac{d}{dt}
\left(
\begin{array}{c}
\drho \\ \dchi
\end{array}
\right)
=
\frac{T(\rho_*)}{\rho_*}
{\cal M}
\left(
\begin{array}{c}
\drho \\ \dchi
\end{array}
\right),
\label{pert} \ee
where
\begin{eqnarray}
{\cal M}&=&
\left(
\begin{array}{cc}
-\nu +3(1-w)/(1-3w)&~~~~~ 2(1-3w) \\
\nu-2/(1-3w) &~~~~~
-2(1+9w)/[3(1+w)(1-3w)]  \end{array}
\right)
\label{mat} \\
\nu &=& \frac{d\ln |T|}{d\ln\rho} \left(\rho_* \right),
\label{aaa}
\end{eqnarray}
and we have employed the relations (\ref{fp}) and $T(\rho) \propto \rho^\nu$.
The eigenvalues of the matrix $\cal M$ are
\be
M_{1,2}=\frac{
7+3w-3\nu (1+w)\pm\sqrt{
24(-3+2\nu)(1+w)+\left[7+3w-3\nu (1+w)\right]^2
}}{6(1+w)}.
\label{eigen} \ee
For $-1<w<1/3$, $0\leq\nu < 3/2$ they both have a positive real part. As we
have assumed  $T(\rho)<0$, the fixed point is stable in this case.
The approach to the fixed-point values depends on the sign of the quantity
under the square root. If this is negative the energy density oscillates
with diminishing amplitude around its fixed-point value.

\section{Tracking solutions}

We will now analyze the case $\nu=3/2$ which lies at the boundary of the stability region discussed above.
We will thus assume that $T=A~\rho^{3/2}$, and that the universe expands and is dominated by non-relativistic matter
(w=0). Then, in the 4d regime,  equation (\ref{zz}) becomes ($\lambda =0$)

\be
2\rho\zeta'={\zeta^2-A\zeta-1\over 3\zeta+A}
\label{track}\ee

We will parameterize the dimension-less coefficient $A$ as $A=\m-{1\over \m}$.
$A$ is determined by the details of the microscopic cross
section that gives rise to this type of energy exchange.
$\m$ running on non-negative real numbers parameterizes all
possible values of $A$.
When we have expansion, $A>0$ means outflow. When we have
contraction $A<0$ means outflow.

The general solution of equation (\ref{track}) is

\be
(\zeta-\m)^{-{2\over \mu}+8\m}~\left(\zeta+{1\over
\mu}\right)^{-2\m+{8\over \m}}=C~\rho^{\m+{1\over \m}}
\label{hr}\ee

Since $H^2=\rho+\chi$ the equation above can be re-written,
 in terms of
$\zeta=H/\sqrt{\rho}$.

Then equation (\ref{zzz}) becomes
\be
{a\over \rho}{d\rho\over da}=-{3\zeta+A\over \zeta}
\ee
and
can be integrated as a function of $a$ with the result
\be
(\zeta-\m)^{2\m^2}~\left(\zeta+{1\over
\mu}\right)^{2}=C'~a^{-(\m^2+1)}
\label{ha}\ee

$\rho(a)$ can be obtained by solving (\ref{hr}) and substituting
into (\ref{ha}).
Finally $\chi(a)=\rho(\zeta^2-1)$.

We will first study a few special cases :

{\bf (i)} $\mu=1$.
Here we obtain \be
\zeta^2=1+C\rho^{1/3}\Rightarrow  H^2=\rho+C~\rho^{4/3}
\sp
 \rho={C'^3\over C^3}{1\over
a^3}\sp \chi={C'^4\over C^3}{1\over
a^4}
\ee
compatible with the absence of energy exchange in this case and consequent
independence of the evolution of $\rho,\chi$.

{\bf (ii)} $\mu=-1/2$.
\be
\zeta=2+C\rho^{1\over 6}\Rightarrow  H^2=\rho(2+C\rho^{1/6})^2
\ee
For large $a$
\be
\rho=\left({2C'\over 5C^2}\right)^{3/2}a^{-15/4}+{\cal
O}\left(a^{-15/2}\right)
\ee
There is a similar solution for $\mu=1/2$ with asymptotic
contraction. The cases $\mu=\pm 2$ are equivalent to the above.

{\bf (iii)} $1/2<\mu$. It corresponds to $-{3\over 2}<A$.
Asymptotically ($a\to \infty$) we obtain the tracking solution

\be
\zeta=\m+\tilde C~\rho^{{\m^2+1\over 2(4\mu^2-1)}}+...\sp
H^2=\m^2 \rho+2\m\tilde C~\rho^{{\m^2+1\over 2(4\m^2-1)}+1}+...
\ee
\be
\rho\sim \tilde C'~ a^{{1\over \m^2}-4}+...\sp
\chi=(\m^2-1)\rho+...
\ee
Here, although the initial conditions for the real $\rho$ and
mirage $\chi$ energy density are arbitrarily different
(parameterized by the independent integration constants $C,C'$),
at late times the scale similarly with the scale factor
\be
\rho\sim \tilde C'~ a^{{1\over \m^2}-4}+...\sp
{\chi\over \rho}=(\m^2-1)+...
\ee
Thus, the dark energy behaves as the visible energy, and such a
mechanism could be used so that bulk energy simulates dark matter.

{\bf (iv) $-2<\mu<0$}  It corresponds to $-{3\over 2}<A$.
For asymptotically small $\rho$ we obtain
\be
\zeta={1\over |\m|}+\tilde C~\rho^{{\m^2+1\over 2(4-\m^2)}}+...\sp
H^2={\rho\over \m^2}+2{\tilde C\over |\m|}~\rho^{{\m^2+1\over 2(4-\m^2)}+1}+...
\ee
On the other hand
\be
\rho\sim \tilde C'~ a^{\m^2-4}+...\sp \chi={1-\m^2\over
\m^2}\rho+...
\ee
Here we have again tracking behavior.

All other ranges have asymptotic $\zeta$ which is negative and thus unphysical.

Finally, in the case of outflow, there is a fixed point in the 5d regime, when  $A^2<9/4$ with
\be
\rho_*={9-4A^2\over 8.81}\sp H_*={A\over 108}\sqrt{9-4A^2\over 2}
\ee
This  is a saddle point

It is interesting to note that a similar tracking behavior has been
observed in matter interacting with the dilaton in \cite{venez}.

\section{Fixed points in the non-linear regime}
\def\rs{\rho_*}
\def\r{\rho}
\def\rt{\tilde \rho}
\def\cs{\chi_*}
\def\hs{H_*}
\def\qs{q_*}
\def\ts{T_*}
\def\l{\lambda}
\def\r{\rho}

We will consider solutions to the non-linear system (\ref{rho1}), (\ref{qq}) with $H=\hs$ constant. The equations
also imply that $\rho=\rs$, $T=\ts$, $\chi=\cs$ are also constant.
We will see that although there may be a leftover cosmological
constant $\lambda$ on the brane, the cosmological acceleration
because of energy inflow, may be much smaller than
$\sqrt{\lambda}$.

From the equations we obtain
\be
\rs^{\pm}={1\over
144(1+3w)}\left[(1-3w)\pm\sqrt{(1-3w)^2+1152(1+3w)(\l-\hs^2)}\right]
\ee
\be
\cs=-{3\over 4}(1+w){\rs}\left[72\rs+1\right]\sp
\ts=-3(1+w)\hs\rs\sp \qs=\hs^2
\ee

Assuming the rate of expansion $\hs$ to be small compared to the
cosmological constant $\l$,  we have the following two possibilities

\bigskip
(i) $\l$ dominates in the square root.  In this case $\rs\simeq \sqrt{\lambda\over 18(1+3w)}$.
There is still space for this approximation to be correct and
$\rs<<1$ so that we are in the 4d period.

\bigskip
(ii) In the opposite case $\rs\simeq {1-3w\over 72(1+3w)}$ and we
can be either in the 4d or the 5d regime.

In either case , energy exchange can mask a leftover brane
cosmological constant

\section{Other accelerating solutions}

We will present here two different families of solutions that are
characteristic in their classes.

\begin{figure}[htb]
\begin{center}
\epsfig{figure=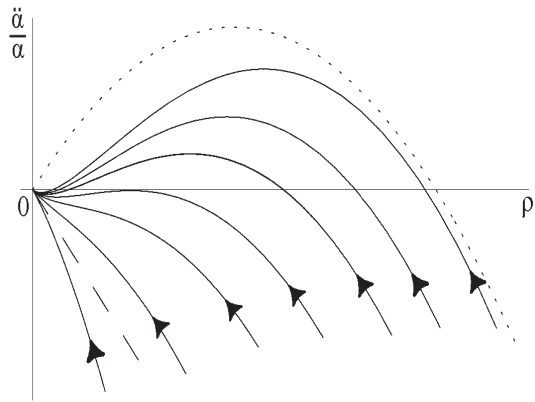, width=200pt,height=150pt} \caption{\label{ea}
\small
\small Outflow, $k=0$, $w=0$,
$\nu=1$. The arrows show the direction of increasing scale factor}
\end{center}
\end{figure}

A global phase portrait of $q\equiv\ddot{a}/a$ with respect to
$\rho$ during expansion in the outflow case for $k=0,\,w=0, \,\nu=1$ is shown
in Figure 1.
All solutions are below the limiting parabola $q<{1-3w\over
2}\rho-36(1+3w)\rho^2$.

One recognizes two families of solutions:
The first have $q<0$ for all values of $\rho$, while the  second
start with a deceleration era for large $\rho$, enter an acceleration era and then
return to deceleration for small enough values of $\rho$.

Solutions corresponding to initial conditions with positive $q$
(always under the limiting parabola shown with the dotted line),
necessarily had a deceleration era in the past,
and are going to end with an eternal deceleration era also.
The straight dashed line represents the standard FRW solution without the effects of
energy exchange.

The global phase portrait of $q\equiv \ddot{a}/a$ with respect to $\rho$ during
expansion for the case $k=0$, $w=0,\nu=1$ is shown in Figure
2. The presence of the limiting parabola as in the outflow case is apparent.
However, new
characteristics appear. For example, $\rho_*^{(-)}$ attracts to eternal
acceleration a whole family of solutions which start their evolution at either
very low or very high densities. There is another family of solutions which
are attracted to acceleration by $\rho_{*}^{(+)}$ and which eventually exit to
a deceleration era.
Finally, there is a family of solutions, near the limiting parabola, which
start with acceleration at very low
densities, and eventually exit to eternal
deceleration, while their density increases
monotonically with time because of the influx.

\begin{figure}[htb]
\begin{center}
\epsfig{figure=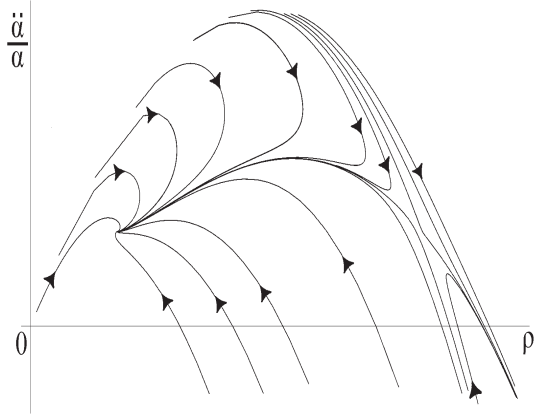, width=200pt,height=150pt} \caption{\label{ea}
\small
\small Influx, $k=0$, $w=0$,
$\nu=1$.}
\end{center}
\end{figure}

For $\nu\neq 1$ one expects a different set of fixed points with
varying behaviors around them.

\section{Conclusions}

An approximate phenomenological analysis of the cosmological
effects of brane-bulk energy exchange indicates the presence of
several interesting phenomena ranging from acceleration (and its
exit) from energy outflow, to inflationary fixed points in the
case of inflow to tracking solutions.

This makes imperative the detailed study of the associated cosmological
dynamics.
I particular the role of the "small-back-reaction approximation"
must be understood. It is expected that this is ok at late times i
the evolution but it breaking at large densities should be
quantified.

A particularly interesting direction  is the holographic
formulation of this problem. This should correspond to the
observable matter, interacting directly to classical four-dimensional
gravity as well as to a "hidden" four-dimensional gauge theory (a
perturbation of N=4 super Yang-Mills). Brane-bulk energy exchange
corresponds to the interaction between observable matter and
"hidden" matter. A quantitative study should clarify several
aspects of the problem.

{\bf Acknowledgement} I would like to thank my collaborators,
 G. Kofinas,  N. Tetradis, T. Tomaras, V. Zarikas
with whom the physics reported here was produced.
I acknowledge discussions with T. Banks, R. Brandeberger,
 R. Brustein, C. Charmousis, J. Cline, L. Cornalba, M. Costa,
 A. Davis, N. Deruelle, N. Kaloper, J. Khoury,  D. Langlois,
 D. Marolf, R.
 Myers, H. Real,
L. Sorbo, D. Steer,     S. Trivedi, C. van der Bruck and G.
Veneziano. I would like to thank the organizers of the 2003 Ahrenshoop
meeting, for giving me the opportunity to lecture,  for hospitality
and an excellent organization of a high quality meeting.
I would like to also thank the organizers of the 2003 Cambridge
workshop on Brane-Worlds as well as those of the String Cosmology
Program in the KITP for hospitality.
This work was partially
supported by European Union under the RTN
contracts HPRN--CT--2000--00122, --00131 and  --00148.


\end{document}